\documentclass[
  aps,
  prl,
  reprint,
  showpacs,
  floatfix,
  superscriptaddress,
  footinbib
]{revtex4-2}

\usepackage[T1]{fontenc}
\usepackage[utf8]{inputenc}
\usepackage[english]{babel}
\usepackage[normalem]{ulem}

\usepackage{amsmath}
\usepackage{mathrsfs} 
\usepackage{lipsum}
\usepackage{amsfonts}
\usepackage{amssymb}
\usepackage{amsthm}
\usepackage{mathtools}
\usepackage{latexsym}
\usepackage{bm}
\usepackage{relsize}
\usepackage{braket}
\usepackage{slashed}
\usepackage{empheq}
\usepackage[makeroom]{cancel}
\usepackage{xfrac} % slanted fractions
\usepackage{upgreek}
\usepackage{ulem} 

\usepackage{graphicx}
\usepackage{tabularx}
\usepackage{multirow}
\usepackage{floatrow}
\usepackage{pgf}

\usepackage{xspace}
\usepackage{hyperref} 
\usepackage[most]{tcolorbox}
\usepackage{comment}
\usepackage{tikz-feynman}
\usepackage{tikz}	
\tikzfeynmanset{compat=1.0.0}

\allowdisplaybreaks

\usepackage{colortbl}
\definecolor{summersky}{cmyk}{0.71,0.33,0,0.5}
\definecolor{flamingo}{cmyk}{0,0.51,0.71,0.5}
\definecolor{rp}{cmyk}{0.2, 1, 0.6, 0}
\definecolor{pacificblue}{cmyk}{0.95,0.3,0, 0.5}
\definecolor{gray60}{cmyk}{0.4,0.4,0,0.8}

\newcolumntype{P}[1]{>{\centering\arraybackslash}p{#1}}

\newcommand{\bfk}{{\mathbf{k}}}

\newcommand{\fig}[1]{ \vcenter{\hbox{\includegraphics[width=0.15\textwidth]{#1.pdf}}} }

\renewcommand{\L}{\mathcal{L}}

\newcommand{\fpi}{f_{\pi}}
\newcommand{\pir}{\pi_{r}}
\newcommand{\pia}{\pi_{a}}
\newcommand{\dpir}{\dot{\pi}_{r}}
\newcommand{\dpia}{\dot{\pi}_{a}}

\newcommand{\fnl}{f_{\rm NL}}

\def\bfk{{\boldsymbol{k}}}

\newcommand{\dd}{\mathrm{d}}

\begin{document}

\title{Primordial non-Gaussianity constraints on dissipative inflation}

 \author{Santiago Ag\"u\'i Salcedo}
\affiliation{Department of Applied Mathematics and Theoretical Physics, University of Cambridge, Wilberforce
Road, Cambridge, CB3 0WA, UK}
 \affiliation{Kavli Institute for Cosmological Physics,
University of Chicago, 5640 South Ellis Ave.,
Chicago, IL 60637, USA}

\author{Thomas Colas}
 \affiliation{Department of Applied Mathematics and Theoretical Physics, University of Cambridge, Wilberforce
Road, Cambridge, CB3 0WA, UK}

\author{Petar Suman}
 \affiliation{Department of Applied Mathematics and Theoretical Physics, University of Cambridge,
Wilberforce Road, Cambridge, CB3 0WA, UK}

\author{Bowei Zhang}
 \affiliation{Department of Applied Mathematics and Theoretical Physics, University of Cambridge,
Wilberforce Road, Cambridge, CB3 0WA, UK}

 \author{James Fergusson}
\affiliation{Department of Applied Mathematics and Theoretical Physics, University of Cambridge, Wilberforce
Road, Cambridge, CB3 0WA, UK}

\author{E. P. S. Shellard}
\affiliation{Department of Applied Mathematics and Theoretical Physics, University of Cambridge, Wilberforce
Road, Cambridge, CB3 0WA, UK}

\begin{abstract}
Dissipative effects appear in many early-Universe scenarios, yet their universal observational signatures and systematic confrontation with data remain largely unexplored. We employ the Open Effective Field Theory of Inflation (Open EFToI) to consistently incorporate dissipative and stochastic effects while preserving scale invariance. Dissipation enhances specific interaction channels of the Goldstone mode, generating distinctive primordial non-Gaussian signatures, beyond those generically produced by standard EFToI. In the weak-dissipation regime, this includes folded bispectrum shapes observationally more favoured than both the equilateral and orthogonal templates. Using the Modal bispectrum pipeline with the Planck CMB data, we obtain the likelihood and derive the first model-independent bounds on early-Universe dissipation. We find a marginalised upper bound on the dissipation scale $\gamma \leq 384\,H$ and a lower bound on the sound speed $c_s \geq 0.38$ at $95\%$ confidence level. The maximum likelihood for best-fit models reveals a degeneracy between $\gamma$ and $c_s$. These results open a model-independent window for probing departures from minimal inflation and discriminating between early-Universe scenarios with stochastic noise and dissipative effects.

\end{abstract}

\maketitle

\paragraph*{Introduction.---} 

Inflation provides a compelling mechanism for generating primordial perturbations, yet its microphysical origin remains unknown. A wide class of early-Universe scenarios, including alternatives to single-field slow-roll inflation, can reproduce the observed nearly scale-invariant power spectrum. This degeneracy raises a central question: how can different realizations of the early Universe be observationally distinguished?

Dissipative and stochastic effects arise in a variety of early-Universe scenarios, including gauge \cite{Anber:2009ua, Peloso:2022ovc}, warm \cite{Berera:1995ie, Berghaus:2025dqi}, and stochastic inflation \cite{Starobinsky:1986fx}. They provide a qualitatively different origin for cosmological inhomogeneities by modifying both the dynamics of fluctuations and their interactions. However, existing analyses typically rely on specific microscopic constructions, making it difficult to identify robust, model-independent signatures. A systematic framework that isolates the universal observational imprint of dissipation is therefore still lacking.

In this Letter, we treat inflationary perturbations as an open system \cite{Salcedo:2025ezu} and extend the effective field theory of adiabatic fluctuations to incorporate dissipative and stochastic effects consistent with scale invariance \cite{Salcedo:2024smn}. This formulation identifies the leading operators that encode energy exchange with additional degrees of freedom while remaining agnostic about their microscopic origin \cite{LopezNacir:2011kk}. We show that dissipation generically enhances specific interaction channels of the Goldstone mode, producing a characteristic pattern of non-Gaussian correlations.

We compute the resulting bispectrum and confront it with current Cosmic Microwave Background (CMB) data \cite{Planck:2019kim}. This yields the first observational bounds on early-Universe dissipation that do not rely on a specific underlying model. These results open a model-independent avenue for testing departures from minimal inflation and sharpen the observational criteria needed to discriminate between competing early-Universe scenarios.\\

%%%%%%%%%%%%%%%%%%%%%%%%%%%%%%%%%%%%%%%%%%%%%%%%%%%%%%%%%%%%%%%%%%%%%

\begin{figure*}[t!]
    \centering
    \includegraphics[width=0.7\linewidth]{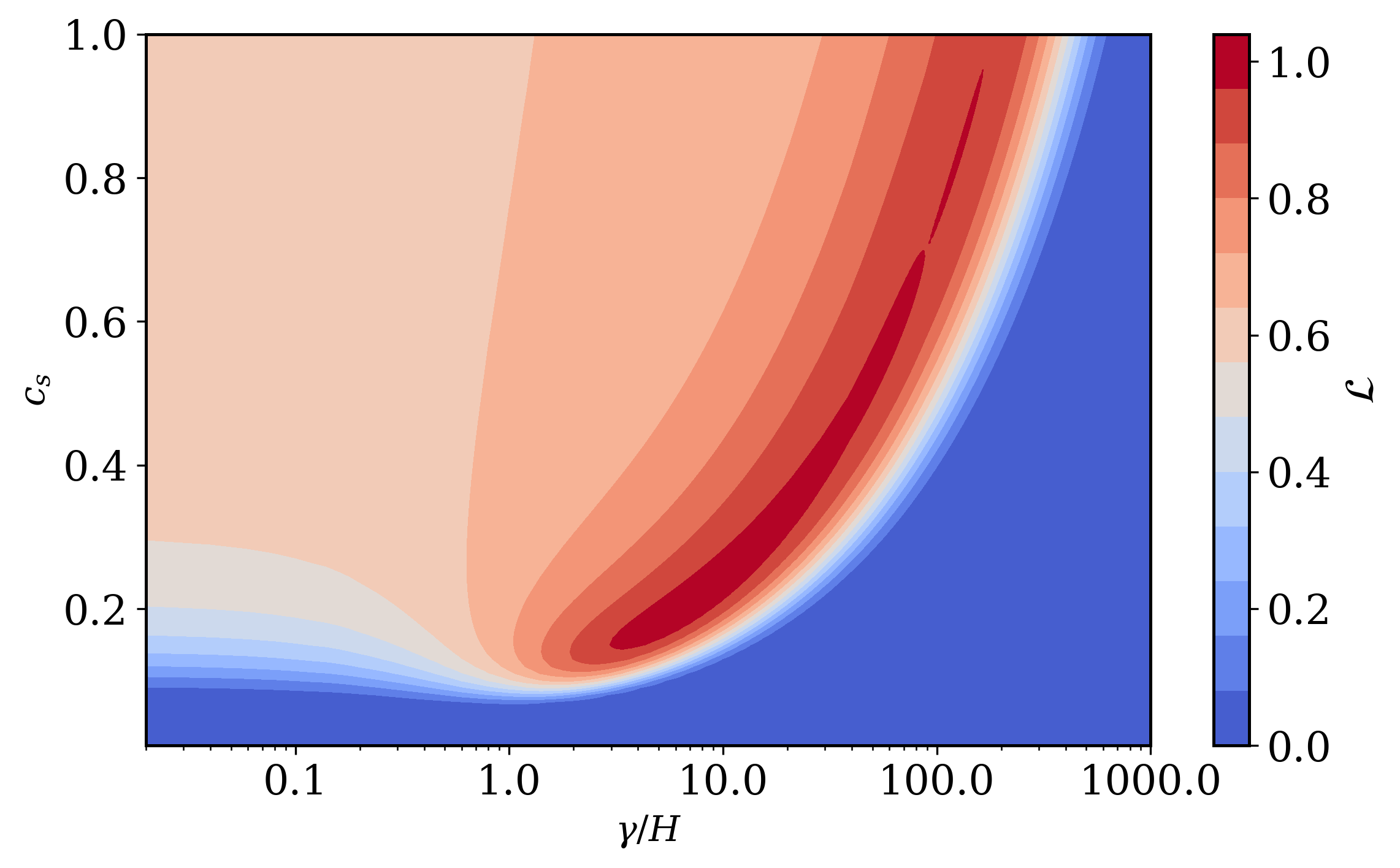}
  \caption{Likelihood \eqref{eqn:-2logL} in the $(\gamma, c_s)$ plane used to constrain the dissipation $\gamma$ and sound speed $c_s$. The likelihood is normalized to unity at its maximum to facilitate visual comparison.}
    \label{fig:likelihood}
\end{figure*}

\paragraph*{A road to dissipative inflation.---} 

In the simplest models of inflation, primordial non-Gaussianities (PNG) are expected to be nearly undetectable \cite{Maldacena:2002vr}. Larger signals can arise in the presence of additional fields \cite{Chen:2009zp,Seery:2005gb,Arkani-Hamed:2015bza} or a small sound speed \cite{Cheung:2007st,Baumann:2011su}. A class of scenarios generates a large population of such fields through particle production mechanisms. One example is axion inflation \cite{Anber:2009ua}, where an axion-like inflaton sources abundant massless gauge bosons. In general the resulting open-system dynamics of the axion are nonlocal, complicating an effective field theory description. Under suitable hierarchies of scales, however, local dissipative dynamics can emerge \cite{Creminelli:2023aly}: an instability produces scalar particles on sub-Hubble scales that subsequently decay into inflaton fluctuations. The resulting dynamics match the Open Effective Field Theory of Inflation (EFToI) \cite{Salcedo:2024smn}. These scenarios are typically far from thermal equilibrium, since particle production occurs out of equilibrium. Nevertheless, imposing the dynamical Kubo-Martin-Schwinger symmetry \cite{Crossley:2015evo} recovers a thermal limit during inflation corresponding to warm inflation \cite{Berera:1995ie}, for which concrete UV realizations have recently been constructed \cite{Berghaus:2019whh,Berghaus:2025dqi}. The Open EFToI therefore provides a versatile parametrization of dissipative inflationary dynamics which we constrain in this Letter.

In \cite{Salcedo:2025ezu}, we built a general framework to study interacting sectors in gravity based on general relativity and the Schwinger-Keldysh formalism \cite{Schwinger:1960qe, Keldysh:1964ud}. The latter provides a systemic description of non-equilibrium and open cosmological systems which relies on a doubled path integral contour, with one direction going forward in time (the ``$+$'' contour), and one direction going backward in time (the ``$-$'' contour) \cite{Colas:2025app}. This doubling enables the derivation, directly from the path integral, of general non-time-ordered correlation functions, including the cosmological correlators that provide the theoretical predictions confronted with CMB temperature anisotropies.

When the slow-roll hierarchy holds, the dynamics of the scalar field fluctuations decouple from the perturbations of the metric and one can safely focus on the evolution of a single-scalar adiabatic degree of freedom $\pi$ \cite{Cheung:2007st, LopezNacir:2011kk, Salcedo:2025ezu}. This quantity relates to the curvature perturbations that seed the CMB anisotropies through $\zeta=-H\pi$, where $H$ is the Hubble parameter. The problem reduces to writing down the theory of a dissipative shift symmetric scalar in a quasi de-Sitter background. Just as any field, $\pi \rightarrow \pi_\pm$ is doubled in the Schwinger-Keldysh contour. It is convenient to analyse the dynamics in the Keldysh basis, in terms of a retarded component $\pir = (\pi_+ + \pi_-)/2$ and an advanced component $\pia = \pi_+ - \pi_-$. In \cite{Salcedo:2024smn}, we derived the most generic local dissipative dynamics of a single-clock perturbation through a functional $S_{\mathrm{eff}} [\pi_r,\pi_a]$. While the conservative theory enjoys invariance under shift symmetry of both $\pi_+$ and $\pi_-$, the presence of an environment generates dissipative and stochastic effects mixing the branches of the path integral. This leads to a symmetry breaking pattern under which $S_{\mathrm{eff}}$ becomes only invariant under the smaller diagonal subgroup: $\mathrm{shift}_+ \times \mathrm{shift}_- 	\rightarrow   \mathrm{shift}_r$.
It implies that only $\pi_r$ non-linearly realises time-translation and boosts. Explicitly, under the time-translation $t \rightarrow t+ \xi$, 
\begin{equation}
\begin{aligned}
    \pir(t,\textbf{x})&\rightarrow \pir(t+\xi,\textbf{x})-\xi,\\
    \pia\left(t,\textbf{x}\right)&\rightarrow\pia\left(t+\xi,\textbf{x}\right). \label{eq:diagonalNLRv0}
\end{aligned}
\end{equation}

Further imposing unitarity of the UV complete theory and locality of the EFT, we wrote in \cite{Salcedo:2024smn} the most generic functional compatible with these principles known as the Open EFT of Inflation. In this Letter, we focus on a subset of this construction controlled by the linear dynamics,
\begin{equation}
    S^{\left(2\right)}_{\mathrm{eff}}= \int \dd^4x \sqrt{-g} \left[\dpir\dpia-\frac{c_{s}^{2}}{a^{2}}\partial_{i}\pir\partial_{i}\pia-\gamma\dpir\pia+i\beta\pia^{2} \right]\,. \bigg. 
\end{equation}
The first two terms correspond to the standard kinetic term with a speed of sound $c_s$, while the last two terms capture the dissipation and noise from the presence of an environment. Due to the non-linearly realised symmetries, cubic operators are generated whose amplitude is controlled by $c_s$ and $\gamma$ \cite{Cheung:2007st, LopezNacir:2011kk, Salcedo:2024smn}, leading to
\begin{align}\label{eq:OEFToIbispectrumv0}
    S^{(3)}_{\mathrm{eff}} &= \int \dd^4 x \sqrt{-g} \bigg\{\frac{c_{s}^{2}-1}{2\fpi^{2}}\Big[\left(\partial_{i}\pir\right)^{2} \dpia +2\dpir\partial_{i}\pir\partial^{i}\pia\nonumber\\
    -&3 \dpir^{2} \dpia\Big] +\frac{\gamma}{2\fpi^{2}}\left[\left(\partial_{i}\pir\right)^{2}-\dpir^{2}\right] \pia \bigg\}\,,
\end{align}
where $\fpi$ is the symmetry breaking scale controlling the magnitude of the kinetic term. 
While the first group of terms in Eq.~\eqref{eq:OEFToIbispectrumv0} are examples of self-interactions of the inflaton described in the closed EFT of inflation \cite{Cheung:2007st}), the second group are an example of non-linear dissipation \cite{LopezNacir:2011kk}. \\

\paragraph*{Dissipative bispectrum.---} 

Among the probes of PNG, the bispectrum of curvature perturbations, $\langle \hat{\zeta}_{\bfk_{1}}\hat{\zeta}_{\bfk_{2}}\hat{\zeta}_{\bfk_{3}}\rangle=\left(2\pi\right)^{3}\delta\left(\bfk_{1}+\bfk_{2}+\bfk_{3}\right)B\left(\bfk_{1},\bfk_{2},\bfk_{3}\right)$, plays a central role. Its physical content becomes more transparent when factorised into a shape function
\begin{align}
    S(x_2,x_3) \equiv (x_2 x_3)^2 \frac{B(k_1, x_2 k_1, x_3 k_1)}{B(k_1,k_1,k_1)}\,,
\end{align}
where $x_2 \equiv k_2/k_1$ and $x_3 \equiv k_3/k_1$ are restricted to the region $\max(x_3, 1- x_3) \leq x_2 \leq 1$, and an amplitude 
\begin{align}\label{eq:fnldef}
    f_{\mathrm{NL}} \equiv \frac{5}{18} \frac{B(k,k,k)}{P^2_\zeta(k)}\,,
\end{align}
The latter is compared to the amplitude of the Gaussian fluctuations encoded in the primordial power spectrum of these primordial fluctuations, $ \langle \zeta_\bfk \zeta_{-\bfk}\rangle = (2\pi)^3 \delta(\bfk + \bfk') P_\zeta(k)$ where $P_\zeta(k) = (2\pi^2/k^3) \Delta^2_\zeta$ and $\Delta^2_{\zeta} = 2.1 \times 10^{-9}$.

Using the Feynman rules of \cite{Salcedo:2024smn}, we compute the tree-level contact bispectrum generated by the cubic operators in Eq.~\eqref{eq:OEFToIbispectrumv0}. Details are given in the Supplementary Material. For a coupling $g$, the bispectrum takes the generic form\begin{align}\label{eq:Bexpref}
    &B(k_1,k_2,k_3) = (-1)^{4-n_d} \frac{H^3}{f_\pi^6} \frac{g}{H^{4-n_d}} \int_{-\infty(1 \pm i \epsilon)}^{0^-} \frac{\dd \eta}{\eta^{4-n_d}} \nonumber \\
    & \,\,\widehat{\mathcal{D}}\left[ G^{K/R}(k_1, \eta) G^{K/R}(k_2, \eta) G^R(k_3, \eta) + 5~\mathrm{perms} \right]\,.
\end{align}
$\widehat{\mathcal{D}}$ is a differential operator schematically representing the $n_d^{\mathrm{th}}$ spatial and temporal derivatives acting on the Green function
\begin{align}\label{eq:GRB2b}
    G^R(k, \eta) = \frac{H^2}{2k^3}  (-k\eta)^{3} \left(\frac{-k\eta}{2} \right)^{ - \nu_\gamma}\Gamma(\nu_\gamma) J_{\nu_\gamma}(-k\eta)\,,
\end{align}
and Keldysh propagator
\begin{align}
    G^K&(k, \eta) = - i\frac{\pi}{4k^3} \beta \left(-2 k\eta \right)^{\nu_\gamma} \Gamma(\nu_\gamma) \nonumber \\
    \bigg[ &Y_{\nu_\gamma}(-k\eta)\int_{-k\eta}^{\infty} \dd z' z^{\prime2 - 2{\nu_\gamma}} J_{\nu_\gamma}(z') J_{\nu_\gamma}(z') \nonumber \\
    -& J_{\nu_\gamma}(-k\eta)\int_{-k\eta}^{\infty} \dd z' z^{\prime2 - 2{\nu_\gamma}} J_{\nu_\gamma}(z') Y_{\nu_\gamma}(z') \bigg]\,, \label{eq:GK}
\end{align} 
with the order parameter $\nu_\gamma = \frac{3}{2} + \frac{\gamma}{2H}$. At low dissipation, the numerical evaluation of Eq.~(\ref{eq:Bexpref}) becomes technically challenging due to the highly oscillatory behaviour of the Bessel functions $J_{\nu_\gamma}$ and $Y_{\nu_\gamma}$, and the slow convergence of the integrand as $\eta \to -\infty$. To overcome this difficulty, we decompose each Bessel function into a linear combination of Hankel functions $H^{(1)}_\nu$ and $H^{(2)}_\nu$, thereby rewriting the original expression as a sum of integrals involving Hankel functions only. Exploiting the single-frequency asymptotic behaviour of Hankel functions in the limit $\eta \to -\infty$, each sub-integral acquires an overall phase factor of the form $e^{iF(k_1,k_2,k_3)\eta}$, where $F$ is a linear combination of $k_1$, $k_2$, and $k_3$. This structure allows us to deform the integration contour into either the upper or lower half of the complex $\eta$-plane, effectively damping the oscillations and significantly improving numerical convergence. A detailed description of the methodology and its numerical implementation for rapidly computing precision bispectra is presented in a companion paper \cite{Longpaper}.

\begin{figure}[t!]
    \centering
    \includegraphics[width=1\linewidth]{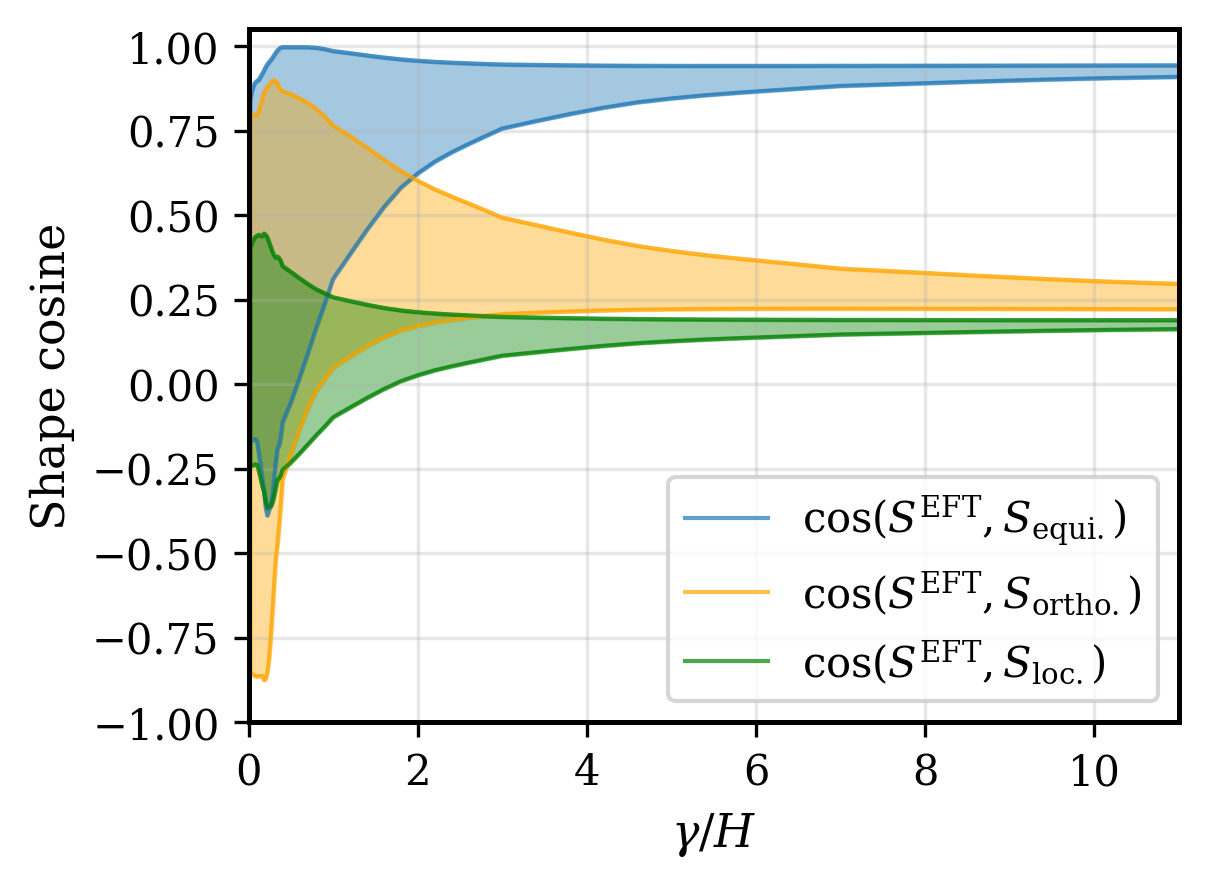}
    \caption{
    Shape correlation of the Open EFToI bispectrum with the standard \textit{equilateral}, \textit{orthogonal} and \textit{local} templates, as used in \textit{Planck} PNG analysis. The shaded regions show the envelope obtained by varying $c_s$ between $0.01$ and $1$. At large dissipation, the signal correlates with the equilateral template at the $90\%$ level.
    At low dissipation, the signal peaks in a strongly folded configuration whose correlations with standard templates depend on $c_s$  and are not generically orthogonal.}
    \label{fig:shapecorrel}
\end{figure}

The bispectra generated by Eq.~\eqref{eq:OEFToIbispectrumv0} share a common phenomenology. At low dissipation, most operators produce a large signal in the folded configuration, $k_{1}=2k_{2}=2k_{3}$. 
This enhancement arises from the classical sourcing of environmental noise \cite{Green:2020whw}. The small dissipation regularizes the signal, distinguishing it from those generated by non-Bunch initial conditions \cite{Chen:2006nt, Planck:2013wtn, Salcedo:2024smn}.
At large dissipation, the dominant contribution instead shifts toward the equilateral configuration, $k_{1}=k_{2}=k_{3}$. In this regime, dissipation damps correlation between modes that freeze at different times.
This transition from strongly folded to equilateral shapes is reflected in the (incomplete) overlap with the standard equilateral, orthogonal, and local templates shown in Fig.~\ref{fig:shapecorrel}.

The predicted amplitude $\fnl^{\rm th}$ of the bispectrum can be estimated using the heuristic argument of \cite{Salcedo:2024smn}. The operators associated with $c_{s}$ in Eq.~\eqref{eq:OEFToIbispectrumv0} yield an approximately constant $\fnl$ at fixed speed of sound, reproducing the EFToI scaling $\fnl\sim c_{s}^{-2}$ in the small-$c_{s}$ limit~\cite{Cheung:2007st}. In contrast, the operators controlled by the dissipation parameter $\gamma$ can generate large PNG at strong dissipation and small $c_{s}$. At large $ \fnl^{\rm th} \gg 1$, it behaves either as
\begin{equation}\label{eq:fnltheo}
    \fnl^{\rm th}\sim \left\{\frac{1}{4c_{s}^{2}}\frac{\gamma}{H};\,\frac{1}{3c_{s}^{2}}\right\}\,.
\end{equation}\\

\paragraph*{Modal analysis.---}
We emphasize that the Open EFToI provides a framework that predicts both the shape and the amplitude of the primordial bispectrum. This contrasts with many conventional PNG calculations, which typically determine the bispectrum shape only up to an arbitrary overall amplitude.

By projecting the primordial bispectrum shape \eqref{eq:Bexpref} through linear transfer functions $T_{\ell}(k)$ onto the CMB signal
\begin{equation}\label{eqn:th_CMB_bispectrum}
    B_{\ell_1 \ell_2 \ell_3}^{(\rm th.)} \sim \int \mathrm{d}r\, r^2 \int \prod_{i=1}^3 \left[ k_i^2 \mathrm{d}k_i j_{\ell_i}(k_i r) \, T_{\ell_i}(k_i) \right] B(k_1,k_2,k_3),
\end{equation}
we directly compare this theoretical bispectrum with the observed CMB bispectrum (up to a linear term, detailed in Supplementary Material) from the Planck 2018 data 
\begin{equation}
    B^{m_1,m_2,m_3 (\text{obs.})}_{\ell_1\ell2\ell_3} = a_{\ell_1m_1}a_{\ell_2m_2}a_{\ell_3m_3},
\end{equation}
to obtain a likelihood function (up to a constant) for the parameters $(c_s, \gamma)$
\begin{equation}\label{eqn:-2logL}
    -2 \log \mathcal{L} = \sum_{\begin{subarray}{c} \ell_1,\ell_2, \ell_3 \\ m_1,m_2, m_3 \end{subarray}} \left(\tilde B^{m_1,m_2,m_3 (\text{obs.})}_{\ell_1,\ell_2, \ell_3} - \tilde B^{m_1,m_2,m_3 (\text{th.})}_{\ell_1,\ell_2, \ell_3} \right)^2\,.
\end{equation}
The bispectrum expressions are modified to produce independent variables with unit variance. Rigorous statistical derivation of the likelihood is provided in the Supplementary Material.

Evaluation of CMB higher-order statistics faces multiple computational challenges, which cannot be addressed by a naive brute force summation. Firstly, a direct numerical evaluation of \eqref{eqn:th_CMB_bispectrum} naively requires a four-dimensional integral over the momenta $k_i$ and the line-of-sight distance $r$. Secondly, direct reconstruction of the observed bispectrum is computationally prohibited by the $\mathcal{O}(\ell_{\rm max}^5)$ scaling. Finally, the same scaling problem prohibits direct evaluation of the sum over multipoles in \eqref{eqn:-2logL}.

The \textit{Modal} estimator \cite{Fergusson:2009nv, Fergusson:2010dm, Fergusson:2014gea, Liguori:2010hx} circumvents prohibitive computational costs by approximating the bispectrum as a sum of separable basis functions, or ``modes''. It utilises a dual-expansion scheme: i) \textit{Primordial expansion}: decomposes the shape in $k$-space; ii) \textit{CMB expansion}: projects the theoretical template in $\ell$-space, and performs similar expansion of the observed bispectrum. For clarity, we express the expansion in an orthonormal basis of mode functions $\{ \mathcal{R}_i (\ell_1,\ell_2,\ell_3)\}$:
\begin{equation}
    \tilde B^{(\text{th.})}
    \sim f_{\rm NL}^{\rm th}\sum_{i=0}^{n} \alpha_i \mathcal{R}_i \quad \text{and} \quad 
    \tilde B^{(\text{obs.})}
    \sim
    \sum_{i=0}^{n} \beta_i \mathcal{R}_i.
\end{equation}
Equipped with the definition of an inner product between the basis functions, the likelihood reduces to a simple modal summation
\begin{equation}\label{eqn:likelihood-modal}
    -2\log \L(c_s, \gamma) = \mathrm{Const.} + \dfrac{f_{\rm NL}^{\rm th^2}}{6}\sum_i \alpha_i^2 - \dfrac{f_{\rm NL}^{\rm th}}{6}\sum_i 2\alpha_i {\beta}_i.
\end{equation}
This greatly speeds up the likelihood estimation and allows for fast and efficient constraints in the multidimensional parameter space of the theory. 
As we predict both the shape and the amplitude, we choose to perform a Bayesian analysis, with the posterior $\mathcal{P}(c_s, \gamma)$ given by
\begin{equation}
    \mathcal{P}(c_s,\gamma) \propto \L(c_s, \gamma |\text{data}) \cdot \Pi(c_s, \gamma).
\end{equation}
The proportionality constant is subsequently absorbed into the normalization of the posterior (such that $\iint \mathcal{P}\dd \gamma \dd c_s = 1$). The posterior depends on the choice of the Bayesian prior, $\Pi(\gamma,c_s)$, which in this first analysis is taken to be uniform across the 2d parameter space; $c_s \in (0.01,1)$ and $\gamma/H \in (0.02, 1000)$.

While $\gamma = 0$ and $0 < c_s < 1$ provide natural borders of the parameter space in one direction, the high $\gamma$ cut-off needs to be suitably chosen. We observe that for high dissipation, the theoretical amplitude \eqref{eq:fnltheo} grows so large that it is highly disfavoured by the weakly non-Gaussian data. This results in a robust computation of confidence regions, independent of the large $\gamma$ extension provided that we keep $\gamma_{\rm max} \gtrsim 500 H$. 

Typical PNG analyses are performed with the $f_{\rm NL}$ parameter allowed to freely vary. This means we can test whether new physically-motivated shapes provide a good fit to the current data \cite{Sohn:2024xzd, Suman:2025tpv, Suman:2025vuf, Philcox:2025bbo}. In this case, the maximum likelihood estimator (MLE) is then
\begin{equation}
    f_{\rm NL}^{\rm obs.} = \dfrac{\sum_i \alpha_i \beta_i}{\sum_i \alpha_i^2}\,,
\end{equation}
and its error $\sigma_{f_{\rm NL}}$ is estimated using full 160 Gaussian simulations.
Both the best-fit value of $\fnl$ and its error are degenerate with the template's amplitude. For example, if we scale the template $S(k_i) \rightarrow \lambda S(k_i)$, the $\fnl$ estimate would scale as $f_{\rm NL}^{\rm obs.} \pm \sigma_{f_{\rm NL}} \rightarrow \lambda^{-1} f_{\rm NL}^{\rm obs.} \pm \lambda^{-1}\sigma_{f_{\rm NL}}$. This motivates us to consider the signal-to-noise ratio (SNR)
\begin{equation}\label{eq:SNR}
    \tilde{\sigma}_{\rm SNR} = \dfrac{ f_{\rm NL}^{\rm obs.}}{\sigma_{f_{\rm NL}}}\,,
\end{equation}
as an amplitude-independent measure of how well a certain theoretical template fits to the data.
\\

\begin{figure}[t]
    \centering
\includegraphics[width=1\linewidth]{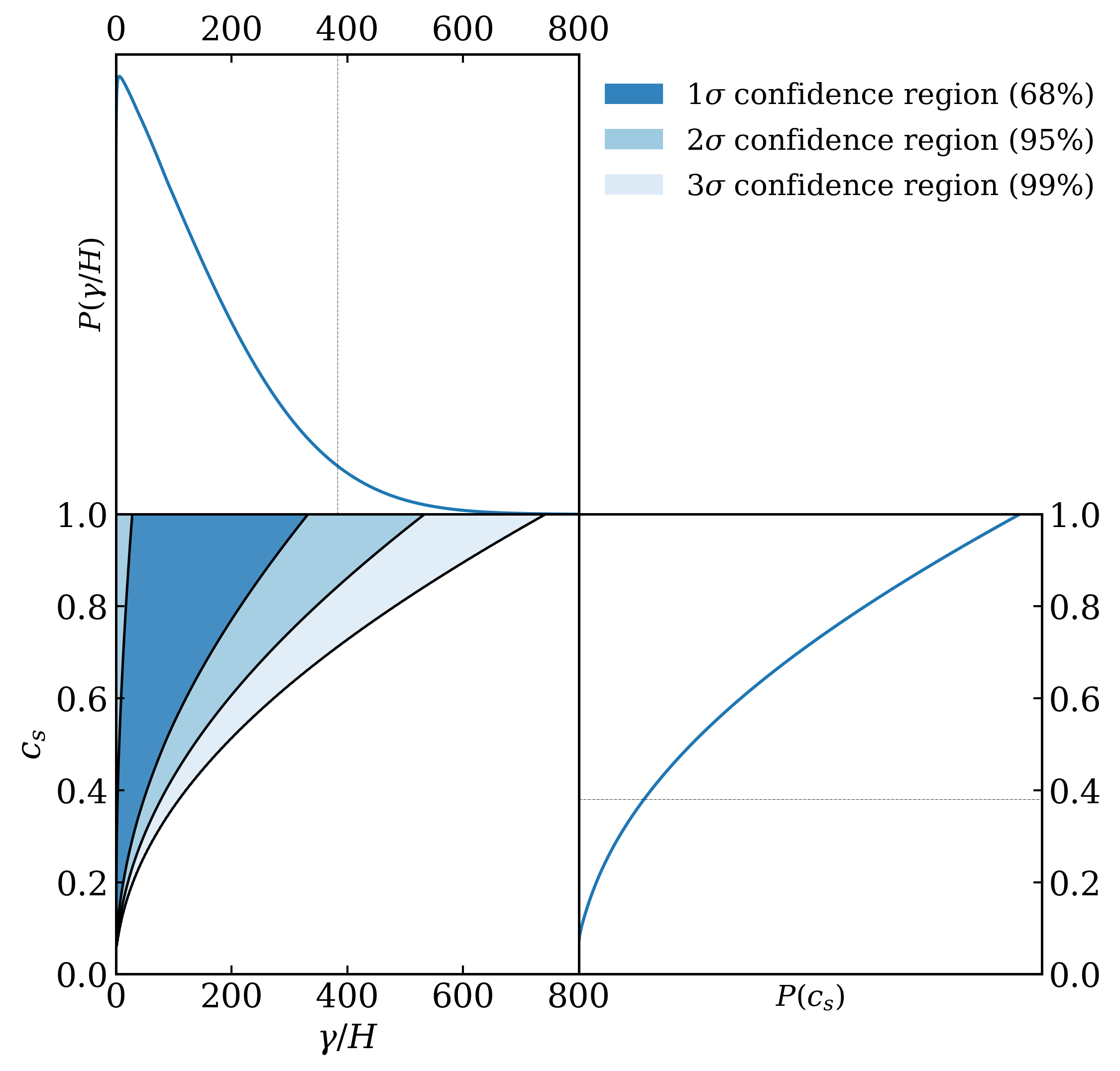}
    \caption{Corner plot of the posterior distributions for the dissipation rate $\gamma$ and sound speed $c_s$, assuming uniform priors. Thin dotted lines mark the $2\sigma$ exclusion limits of the marginalized posteriors, yielding $\gamma < 383.6H$ and $c_s > 0.38$ at 95\% confidence level.
    }
    \label{fig:triangular plots}
\end{figure}

%%%%%%%%%%%%%%%%%%%%%%%%

\paragraph*{Main results.---} 
We now present the main results from the Modal comparison with the Planck CMB data: the likelihood and posterior distributions, and the signal-to-noise ratio (SNR). 

%%%%%%%%%%%%%%%%%

Our Modal likelihood analysis \eqref{eqn:-2logL} exploits the full bispectrum information, including both shape and amplitude, which distinguishes it from previous shape-only analyses of PNG (see \cite{Zhang:2025nyb}). 
The likelihood map shown in Fig.~\ref{fig:likelihood} reflects the interplay between dissipation $\gamma$, which primarily determines the bispectrum shape, and the sound speed $c_s$, which largely controls the amplitude of $\fnl^{\rm th}$. 
The maximum likelihood point corresponds to $(c_s, \gamma/H)_{\rm best-fit} = (0.2, 6.6)$. However, the maximum-likelihood region (dark red) extends from low speed of sound ($c_s\sim0.2$) with moderate dissipation ($\gamma\sim5H$) up to moderate speed of sound ($c_s\sim0.6$) with large dissipation ($\gamma\sim100H$). This degeneracy is a consequence of the shape barely changing for $\gamma\geq10H$ (see Fig.~\ref{fig:shapecorrel}), while $c_{s}$ and $\gamma/H$ play a degenerate role in the expression of $\fnl^{\rm th}$ given in Eq.~\eqref{eq:fnltheo}. A sharp transition from high to low likelihood (red to blue) delineates a disfavoured region with high amplitude $\fnl^{\rm th}$ caused by very small $c_s$ or large $\gamma$. Physically, this region corresponds to an overproduction of PNG not consistent with  Planck measurements. This illustrates how bispectrum measurements constrain the parameter space of dissipative inflation.

%%%%%%%%%%%%%%%%%%%%%%%%%%%%%%%
We can use the likelihood to derive constraints on the parameter space. The posterior distributions of $c_s$ and $\gamma$ shown in Fig.~\ref{fig:triangular plots} are obtained assuming uniform priors on both parameters. The marginalized posteriors imply an upper bound on the dissipation, $\gamma \leq 384H$, and a lower bound on the speed of sound, $c_s \geq 0.38$ at $95\%$ confidence level. These results generalize the lower bound on $c_s$ obtained by Planck for cold single-field inflation \cite{Planck:2019kim}. As in that case, these constraints are primarily driven by the PNG enhancements in Eq.~\eqref{eq:fnltheo}, with increasing $\gamma$ or decreasing $c_s$ leading to $\fnl^{\rm th}$ exceeding the observed value.
The degeneracy visible in the two-dimensional posterior of Fig.~\ref{fig:triangular plots} reflects the likelihood structure shown in Fig.~\ref{fig:likelihood} plotted the $\gamma$ on a log scale. The point $\gamma=0$, $c_s=1$ lies close to the $1\sigma$ contour and is therefore consistent with current observations. One may further ask whether large values of $\gamma$ remain within the regime of validity of the EFT. As shown in the Supplementary Material, the strong-coupling regime defined by $\fnl^{\rm th}\Delta_\zeta\sim1$ lies far from the preferred region. It corresponds to $\fnl^{\rm th}\gg\fnl^{\rm obs}$ and is therefore strongly disfavoured, while the $1\sigma$, $2\sigma$, and $3\sigma$ regions remain well within the EFT regime of validity.

\begin{figure*}[t]
    \centering
    \includegraphics[width=0.65\textwidth]{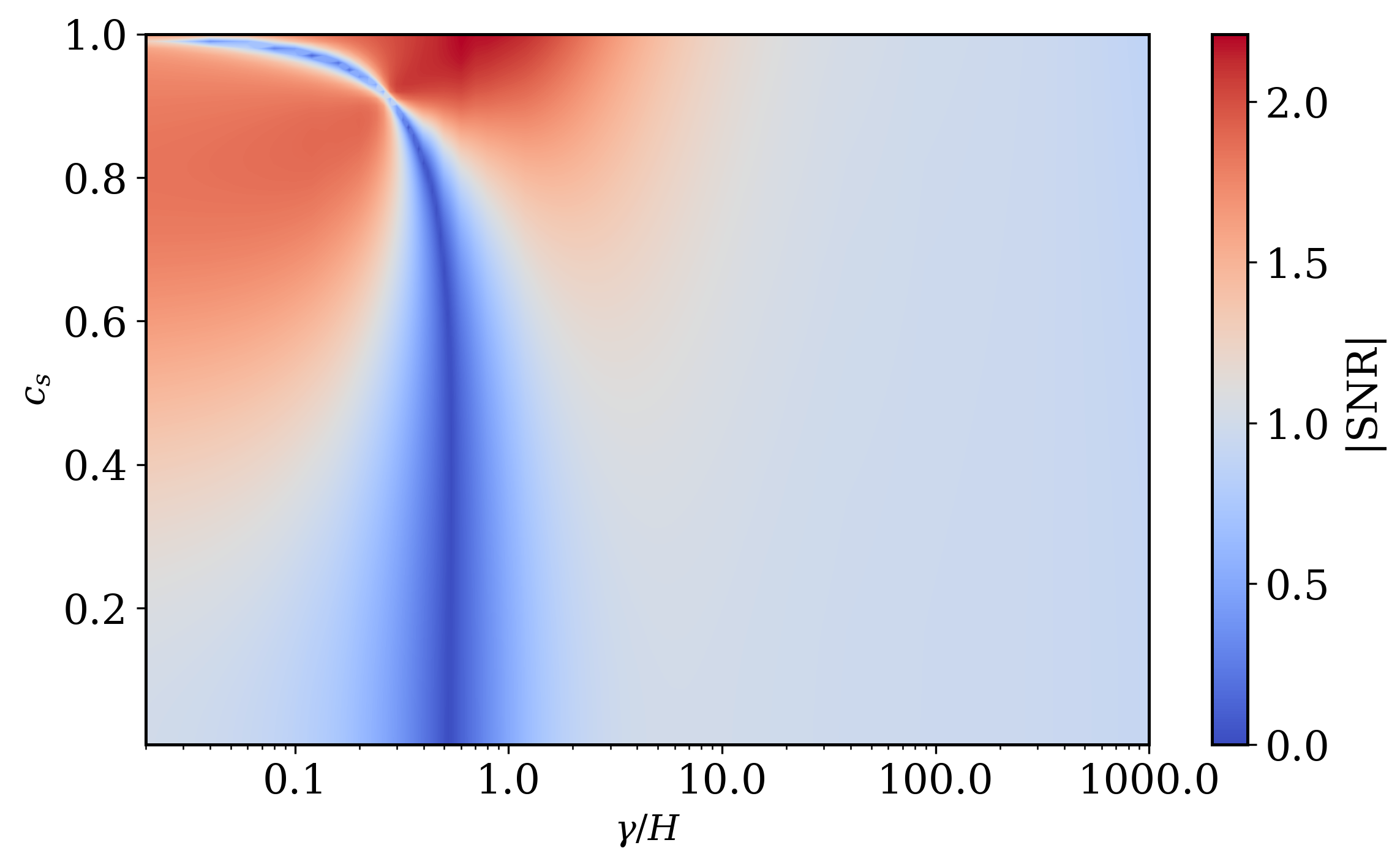}
    \caption{
    Absolute signal-to-noise ratio (SNR) as a function of the sound speed $c_s$ and dissipation $\gamma$, computed from the shape of the predicted bispectrum. The SNR decreases for increasing $\gamma$ and decreasing $c_s$. The differences with the likelihood in Fig.~\ref{fig:likelihood} reflect that Eq.~\eqref{eq:SNR} depends only on the bispectrum shape, while the likelihood Eq.~\eqref{eqn:likelihood-modal} also incorporates the amplitude through $f_{\rm NL}^{\rm th}$. }
    \label{fig:SNR}
\end{figure*}

To assess how well the bispectrum shapes predicted by the Open EFToI fit the CMB data, we plot the signal-to-noise ratio (\refeq{eq:SNR}) as a function of the dissipation and sound speed in Fig.~\ref{fig:SNR}. The signal-to-noise ratio compares the CMB best-fit value of $\fnl^{\rm obs}$ to its uncertainty $\sigma(\fnl^{\rm obs})$ disregarding the theoretical prediction for $f_{\mathrm{NL}}^{\mathrm{th}}$. 
This explains the strong apparent differences between Figs.~\ref{fig:likelihood} and \ref{fig:SNR}. 
The largest SNR (dark red region) occurs near $\gamma\approx H$ and $c_s\approx1$, where the bispectrum features a large negative folded component together with a small but non-vanishing equilateral contribution (Fig.~\ref{fig:shapecorrel}). Increasing the dissipation or decreasing the speed of sound ($\gamma>10$, $c_s<0.4$) leads to a regime dominated by a large equilateral signal with a subdominant folded component, for which the absolute SNR saturates at $\sim1$ (light blue region). Two regions instead yield a nearly vanishing SNR (dark blue). The upper branch ($c_s>0.9$) corresponds to a negligible equilateral signal, while the lower branch ($c_s<0.9$) suppresses the folded contribution; both cases lead to strongly disfavoured correlations with the observed bispectrum $B^{\mathrm{obs}}$. \\

%%%%%%%%%%%%%%%%%%%%%%%%%%%%%%%%%%%%%%%%%%%%%%%%%%%%%%%%%%%%%%%%%%%%%%%

\paragraph*{Physical implication and outlook.---} 
The results presented here initiate a precision study of the primordial bispectrum shape for the Open EFToI. Our computation involves no factorized or approximate templates. Instead, we developed a numerical code to evaluate the bispectrum~\eqref{eq:Bexpref} and extract its predicted amplitudes and shapes. 
We find that stochastic and dissipative effects during inflation can produce distinctive non-Gaussian signatures that cannot be mimicked by single-field cold inflation, notably by generating a sizeable folded signals. In particular, dissipative dynamics predict a large negative folded signal, which is favoured by current \textit{Planck} data relative to the equilateral template ~\cite{Planck:2019kim} by nearly a factor of four at its maximum value shown in Fig.~\ref{fig:SNR}.

These results provide the first model-independent observational analysis of dissipative dynamics in the early universe. Next-generation CMB surveys will be able to directly test the origin of inflationary fluctuations, whether classical or quantum, through these primordial non-Gaussian signals. In particular, as the noise levels decrease and
progressively smaller scales become accessible, we expect tighter posteriors through the increasingly constrained folded signal. Beyond the CMB, large-scale structure surveys are emerging as a powerful probe of PNG~\cite{Chudaykin:2025vdh}. The growing precision of these surveys will be instrumental in tightening constraints on dissipative dynamics during inflation. \\

\paragraph*{Acknowledgments.---} We thank William R. Coulton, Mikhail M. Ivanov, Steven Gratton, Austin Joyce, Enrico Pajer, Oliver H.E. Philcox, Frank Qu, B. Salehian, Wuhyun Sohn, Xi Tong and Dong-Gang Wang for the insightful discussions. B.Z. acknowledges Xi Tong's helpful comments on the numerical calculation of open EFT bispectra, which constitute the foundation of this project. This work has been supported by STFC consolidated grants ST/P000673/1 and ST/X001113/1, and the STFC DiRAC HPC grants ST/P002307/1, ST/R002452/1 and ST/R00689X/1. J.F., P.S. and B.Z. are supported by STFC grant ST/X508287/1.
 S.A.S. was supported by a Harding Distinguished Postgraduate Scholarship. This work was supported by the Kavli Institute for Cosmological Physics at the University of Chicago, and the Kavli Foundation.

\bibliographystyle{JHEP}
\bibliography{biblio}

%%%%%%%%%%%%%%%%%%%%%%%%%%%%%%
\clearpage

\appendix

\begin{widetext}

\section{Supplementary Material}

This Supplementary Material provides technical details underlying the results of the main text. We first review the construction of the EFToI, then extend it to include dissipative and stochastic effects, and finally present the Modal pipeline and the CMB bispectrum.

%%%%%%%%%%%%%%%%%%%%%%%%%%%%%%%%%%%%%%%%%%%%%%%%%%%%%%%%%%%%%%%%%%%%%%%%%%%%%%%%%%%%%%%%%%%%%%%%%%%%%%%%%%%%%%%%%%%%

\subsection{The Effective Field Theory of Inflation}

This note has mainly focused on the Open EFT of Inflation \cite{Salcedo:2024smn}. The development of this paradigm came about from the application of several developments in the field of theoretical cosmology in the last twenty years. 
One of the instrumental advancements came from the EFT of Inflation \cite{Cheung:2007st} later on extended in \cite{LopezNacir:2011kk} to include dissipative and stochastic effects. Here, 
we provide 
a simplified version of the pathway taken in this work to constraint dissipative dynamics in the early universe.

\subsubsection{Unitary gauge} 

Single-field inflation contains three types of degrees of freedom. The first are scalar degrees of freedom. These are the metric fluctuations (constraints) and the fluctuations of the scalar driving inflation (dynamical). The second type are vector fluctuations of the metric degrees of freedom. These redshift so quickly that can be neglected. The third type are the gravitational waves, which correspond to transverse traceless perturbations of the metric. In principle, the Lagrangian for this scenario of inflation only depends on the inflaton field and the metric, $\mathcal{L}=\mathcal{L}\left(\Phi,g_{\mu\nu}\right)$. 

The theory enjoys diffeomorphism invariance. This invariance translates to the freedom of choosing a clock, that is, choosing how we tell time. The choice relevant here is considering the inflaton to be our clock, fixing it to its background value $\Phi\left(t,\textbf{x}\right)=\bar{\Phi}\left(t\right)$.
Plugging this choice back into the Lagrangian implies that several operators take a new form, such as
\begin{equation}
    \partial_{\mu}\Phi\rightarrow \dot{\bar{\Phi}}\left(t\right)\delta_{\mu}^{0}\;,\qquad \; \partial_{\mu}\Phi\partial^{\mu}\Phi = - \dot{\bar{\Phi}}^{2}g^{00}\;,\;...
\end{equation}
which means that the Lagrangian in the unitary gauge can be expanded entirely as a function of time, the metric, a normal vector $\partial_{\mu}\Phi\sim n_{\mu}$ and their derivatives:
\begin{equation}
    \mathcal{L}=\mathcal{L}\left(g_{\mu\nu},n_{\mu},\partial g_{\mu\nu},\nabla_{\mu}n_{\nu},...; t\right).
\end{equation}

Another way to see this change is that the choice of a clock spontaneously breaks time reparametrisation, which means we have to write a Lagrangian that is invariant only under spatial diffeomorphisms. The first terms in the Lagrangian would correspond to Einstein gravity, the kinetic term and the potential of the inflaton:
\begin{equation}
    \mathcal{L}=\frac{M_{\rm pl}^{2}}{2}R-\frac{1}{2}\left(\partial\Phi\right)^{2}-V\left(\Phi\right)+...
\end{equation}
which in unitary gauge become:
\begin{equation}
    \mathcal{L}=\frac{M_{\rm pl}^{2}}{2}R-c\left(t\right)g^{00}-\Lambda\left(t\right)+\sum_{n=2}^{+\infty}\frac{M_{n}^{4}\left(t\right)}{n!}\left(1+g^{00}\right)^{n}+...
\end{equation}
The first three terms of the Lagrangian are the universal part of the EFToI. The rest of the terms correspond to interactions between the inflaton and the metric. We have made the tower of operators in $1+g^{00}$ manifest as it will play a role when we discuss decoupling between gravity and the scalar dynamical degree of freedom. The time dependence of $c(t)$ and $\Lambda(t)$ is fixed entirely in terms of the background equations of motion:
\begin{align}
    3M_{\rm pl}^{2}H^{2}\left(t\right)&=c\left(t\right)+\Lambda\left(t\right)\,,\\
    2M_{\rm pl}^{2}\dot{H}\left(t\right)&=-2c\left(t\right).
\end{align}
The compatibility of the background equations means that $c\left(t\right)$ and $\Lambda(t)$ are not independent. They satisfy a continuity-like equation:
\begin{equation}\label{eq:continuity}
    \dot{c}\left(t\right)+\dot{\Lambda}\left(t\right)+6Hc\left(t\right)=0.
\end{equation}
If we were to recast $c(t)$ and $\Lambda(t)$ in terms of the inflaton field we would see that this equation is the same as the Klein-Gordon equation for the $\bar{\Phi}(t)$ profile. 

\subsubsection{Decoupling limit}

One of the limitations of the unitary gauge is that it is not straightforward to extract the relevant dynamics for inflationary observables. In particular, the scalar degree of freedom $\zeta$, naively associated with the fluctuations of the inflaton, is not in the trace fluctuations of the spatial metric.
It is possible to derive the effective quadratic action for $\zeta$, as done in \cite{Maldacena:2002vr}. This is done by solving for the constrained degrees of freedom of the metric in terms of $\zeta$. An alternative approach is provided by the Stueckelberg trick. This is the introduction of a new field $\pi$ that transforms non-linearly under time reparametrisations such that under a transformation $t\rightarrow t+\xi\left(t,\textbf{x}\right)$, 
the field transforms as
\begin{equation}\label{eq:NLRtime}
    \pi\left(t,\textbf{x}\right)\rightarrow \pi\left(t+\xi,\textbf{x}\right)-\xi\left(t,\textbf{x}\right).
\end{equation}
It implies that, at linear order, we can reabsorb $\zeta$ into $\pi$ through
\begin{equation}
    -H\left(t\right)\pi\left(t,\textbf{x}\right)=\zeta\left(t,\textbf{x}\right)\, .
\end{equation}
Therefore, studying the dynamics of $\pi$ is akin to consider the physics of $\zeta$. We can hence focus on the interactions of $\pi$ with other metrics of degrees of freedom to study in what regime we can neglect any mixing with gravity. The introduction of $\pi$ rewrites the universal part of the EFToI as
\begin{equation}
    \mathcal{L}=\frac{M_{\rm pl}^{2}}{2}R-c\left(t+\pi\right)g^{\mu\nu}\partial_{\mu}\left(t+\pi\right)\partial_{\nu}\left(t+\pi\right)-\Lambda\left(t+\pi\right)+\frac{M_{2}^{4}\left(t\right)}{2}\left[1+g^{\mu\nu}\partial_{\mu}\left(t+\pi\right)\partial_{\nu}\left(t+\pi\right)\right]^{2}+...
\end{equation}
The tadpole generated by varying this Lagrangian with respect to $\pi$ leads to the continuity equation \eqref{eq:continuity}. The linearised Einstein equations allow to solve for the constrained degrees of freedom of the metric in terms of $\pi$:
\begin{equation}
    \delta g_{00}=-2\phi\,,\quad g_{0i}=a\left(t\right)\partial_{i}F\,,
\end{equation}
such that
\begin{equation}
    \phi=\frac{c(t)\pi}{HM_{\rm pl}^{2}}\,,\quad \nabla^{2}F=a(t)\left[\frac{\epsilon^{2}H^{2}}{c_{s}^{2}}\pi-\frac{\epsilon H}{c_{s}}\dot{\pi}\right], 
\end{equation}
where $\epsilon$ is the first slow-roll parameter and $c_{s} < 1$ for $M_{2}^{4}\neq0$. The mixing with gravity is dictated by couplings of the form $\phi\,\dot{\pi}$ or $F\, \nabla^{2}\pi$. Therefore, decoupling between the scalar fluctuations represented by $\pi$ and the metric is given by when the kinetic term of $\pi$ dominates over this mixing:
\begin{equation}
    \dot{\pi}^{2}\gg\phi \, \dot{\pi} \quad \rightarrow \quad \dot{\pi}\gg \epsilon H\pi. 
\end{equation}
This means that the characteristic frequency of the fluctuations has to be well above $\epsilon H$. As this frequency is set by $H$ we have that as long as the slow-roll parameters are small it is safe to neglect any mixing with gravity for the field $\pi$. This is an instrumental simplification, rather than having to consider the complicated multivariable Lagrangian of the EFToI, it is sufficient to consider the theory of $\pi$.

\subsubsection{Lagrangian for the pseudo-Goldstone $\pi$}

We have established that single field inflation takes place at high enough energies to decouple from the metric. Therefore, we can focus solely on $\pi$ and build an EFT that accounts for its non-linearity under time reparametrisation. For this we can take a set of invariant objects and build the Lagrangian for $\pi$. The invariant objects are:
\begin{equation}
    f(t+\pi)\,,\quad \partial_{\mu}\left(t+\pi\right)\partial^{\mu}\left(t+\pi\right)\,,\quad \text{h.d.}
\end{equation}
where $\text{h.d.}$ stands for higher derivative operators. The resulting Lagrangian is just a sum over these invariant blocks, which equivalently corresponds to the EFToI Lagrangian with all the metric perturbations set to zero:
\begin{equation}
     \mathcal{L}=-c\left(t+\pi\right)g^{\mu\nu}\partial_{\mu}\left(t+\pi\right)\partial_{\nu}\left(t+\pi\right)-\Lambda\left(t+\pi\right)+\frac{M_{2}^{4}\left(t\right)}{2}\left[1+g^{\mu\nu}\partial_{\mu}\left(t+\pi\right)\partial_{\nu}\left(t+\pi\right)\right]^{2}+...
\end{equation}

To study the phenomenology of this Lagrangian we need to expand it in powers of $\pi$. The linear order expansion corresponds to the tadpole:
\begin{equation}
    \mathcal{L}=\left[\dot{c}\left(t\right)+\dot{\Lambda}\left(t\right)+6Hc\left(t\right)\right]\pi+...
\end{equation}
which, as mentioned before, vanishes. The vanishing of the tadpole is essential: otherwise the expansion in $\pi$ is not performed around a solution of the background equations of motion, and therefore does not describe the fluctuations associated with $\zeta$. The expansion to quadratic order is given by
\begin{equation}
    \mathcal{L}=-c\left(t\right)g^{\mu\nu}\partial_{\mu}\pi\partial_{\nu}\pi+2M_{2}^{4}\left(t\right)\dot{\pi}^{2}+...
\end{equation}
It is convenient to normalized $\pi$ to be in units of energy. The quadratic Lagrangian takes the form
\begin{equation}\label{eq:quadraticdec}
    \mathcal{L}=\frac{1}{2}\left[\dot{\pi}_{c}^{2}-\frac{c_{s}^{2}}{a^{2}\left(t\right)}\left(\partial_{i}\pi_{c}\right)^{2}\right]=\frac{\fpi^{4}}{c_{s}^{3}}\frac{1}{2}\left[\dot{\pi}^{2}-\frac{c_{s}^{2}}{a^{2}\left(t\right)}\left(\partial_{i}\pi\right)^{2}\right],
\end{equation}
where
\begin{equation}
    \pi=\frac{c_{s}^{3/2}\pi_{c}}{\fpi^{2}}\,,\quad \fpi^{4}=2c_{s}\epsilon H^{2}M_{\rm pl}^{2}\,,\quad \frac{1}{c^{2}_{s}}=1+\frac{2M_{2}^{4}\left(t\right)}{c(t)}.
\end{equation}
The quadratic Lagrangian is sufficient to compute the power spectrum of $\zeta$,
\begin{equation}\label{eq:PkEFToI}
    \langle\zeta^{2}\rangle=\frac{c_{s}^{3}H^{2}}{\fpi^{4}}\langle\pi^{2}_{c}\rangle=\frac{H^{2}}{\fpi^{4}}\frac{H^{2}}{2k^{3}}=\frac{\Delta_{\zeta}}{k^{3}}\quad \rightarrow \quad  \Delta_{\zeta}=\frac{1}{2}\frac{H^{4}}{\fpi^{4}}\,.
\end{equation}
As $\Delta_{\zeta}$ is a very well constrained number by CMB observations it is possible to estimate in this conservative case $\fpi\sim 60H$. 

\subsubsection{Primordial non-Gaussianities} 

The main object of study of this work is the bispectrum of $\zeta$. This may receive many contributions. In the paradigm of single field inflation the dominant contribution is coming from the self interactions of the scalar fluctuations. The leading contributions to the bispectrum are given by cubic self interactions. This means we need to expand the EFToI Lagrangian to cubic order,
\begin{equation}\label{eq:EFToIcubic}
    \mathcal{L}=...+C_{\dot{\pi}^{3}}\dot{\pi}_{c}^{3}+\frac{c_{s}^{3/2}}{f_{\pi}^{2}}\left(\frac{1}{c_{s}^{2}}-1\right)\frac{\dot{\pi}_{c}(\partial_{i}\pi_{c})^{2}}{a^{2}(t)}+...
\end{equation}
where $C_{\dot{\pi}^{3}}$ depends on $M_{3}^{4}$ and $c_{s}$. It is at this level of the expansion that the non-linear realisation of time reparametrisations becomes instrumental. This non-linearity relates different orders in the expansion of $\mathcal{L}$. This means that placing a constraint on the bispectrum is equivalent to placing a constraint on $c_{s}$, as it is one of the two parameters that controls the bispectrum. The primordial bispectrum for $\zeta$ is proportional to the three point function of $\pi_{c}$:
\begin{equation}
    \langle\hat{\zeta}_{\bfk_{1}}\hat{\zeta}_{\bfk_{2}}\hat{\zeta}_{\bfk_{3}}\rangle=-\frac{H^{3}c_{s}^{9/2}}{\fpi^{6}}\langle\hat{\pi}_{c,\bfk_{1}}\hat{\pi}_{c,\bfk_{2}}\hat{\pi}_{c,\bfk_{3}}\rangle=\left(2\pi\right)^{3}\delta\left(\bfk_{1}+\bfk_{2}+\bfk_{3}\right)B\left(\bfk_{1},\bfk_{2},\bfk_{3}\right).
\end{equation}
Inflation is translation, rotation and (approximately) scale invariant. This means that the function $B\left(\bfk_{1},\bfk_{2},\bfk_{3}\right)$ can only depend on the length of the momenta $k_{a}=|\bfk_{a}|$. Furthermore, it implies that it has a definite scaling under scale transformations
\begin{equation}
    B\left(\lambda k_{1},\lambda k_{2},\lambda k_{3}\right)=\frac{1}{\lambda^{6}}B\left(k_{1},k_{2},k_{3}\right).
\end{equation}
These properties invite the parametrization
\begin{equation}\label{eq:definitionfnlshape}
    B\left(k_{1},k_{2},k_{3}\right)=\frac{18}{5}\fnl \frac{S\left(x_{2},x_{3}\right)}{\left(k_{1}k_{2}k_{3}\right)^{2}}P_{\zeta}^{2}\,,\quad \text{where}\quad x_{a} \equiv \frac{k_{a}}{k_{1}},
\end{equation}
with the overall amplitude of the bispectrum (compared to the power spectrum's  $P_{\zeta} = \Delta_\zeta/k^3$) is given by the number $\fnl$. The function $S\left(x_{2},x_{3}\right)$ encodes the kinematic dependence of the bispectrum. It is dubbed the shape function. The bispectrum of the EFToI is given (at leading order) by the sum of two diagrams:
\begin{equation}
    B=\fig{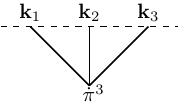}+\fig{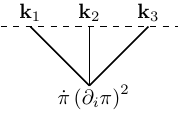}
\end{equation}
The outcome of these two Feynman diagrams are rational functions of the momenta:
\begin{equation}
    B=\frac{\text{Poly}_{6}\left(k_{1},k_{2},k_{3}\right)}{k_{1}^{2}k_{2}^{2}k_{3}^{2}\left(k_{1}+k_{2}+k_{3}\right)^{3}}\,,
\end{equation}
which makes a direct study of the induced PNG computationally expensive. To overcome this, factorised templates have become widely used. In particular, two templates have been tailored to characterise these two shapes. The first is the equilateral template:
\begin{equation}\label{eq:eqShape}
    S^{\rm eq}\left(x_{2},x_{3}\right)=-\left(x_2^2+x_3^2+x_2^2x_3^2\right)-2x_2x_3+\left(x_2^2x_3+x_2x_3^2+\frac{x_3}{x_2}+\frac{x_2}{x_3}+\frac{1}{x_2}+\frac{1}{x_3}\right)\,.
\end{equation}
This template very well approximates the bispectrum sourced by $\dot{\pi}_{c}\left(\partial_{i}\pi_{c}\right)^{2}$ and to some degree the one sourced by $\dot{\pi}^{3}$. To account for this difference the orthogonal shape was introduced:
\begin{equation}\label{eq:orthShape}
    S^{\rm orth}\left(x_{2},x_{3}\right)=-3\left(x_2^2+x_3^2+x_2^2x_3^2\right)-8x_2x_3+3\left(x_2^2 x_3+x_2x_3^2+\frac{x_3}{x_2}+\frac{x_2}{x_3}+\frac{1}{x_2}+\frac{1}{x_3}\right).
\end{equation}

The amplitudes of these shapes are function of the speed of sound and $C_{\dot{\pi}^{3}}$. In particular, large PNG are associated with a small speed of sound as
\begin{equation}
    \lim_{c_{s}^{2}\ll1}\fnl\propto\frac{1}{c^{2}_{s}}\,.
\end{equation}
The amplitude of these shapes has been constrained using Planck 2018 data \cite{Planck:2019kim}, giving
\begin{equation}
    \fnl^{\rm eq}=-26\pm47\,,\quad \fnl^{\rm orth}=-38\pm24\,,
\end{equation}
which has been translated to a bound on the speed of sound $c_{s}\geq0.024$.

\subsubsection{Strong coupling scale}

One of the most important aspects of any EFT is the regime of validity. At sufficiently high energies, other degrees of freedom come at play, or strong coupling will become relevant, and the EFT description stop being valid. Therefore, it is instrumental to derive the scales at which the validity of the EFT breaks down. In the case of the EFToI, one can consider the energy scale $\Lambda_{\ast}$ at which perturbation theory breaks down, that is when the EFT becomes strongly coupled. This was first addressed in \cite{Baumann:2011su}. Here we do a qualitative derivation of the strong coupling scale in the small speed of sound scenario. We first express $\fnl$ in terms of the strong coupling scale and the symmetry breaking scale and then we estimate $\fnl$ based on the argument of \cite{Baumann:2011su}. Perturbation theory assumes that interactions, given in $\mathcal{L}_{3}$ are subleading to the quadratic theory $\mathcal{L}_{2}$. This means that the EFT is weakly coupled as long as $\left|\mathcal{L}_{3}\right|\ll\left|\mathcal{L}_{2}\right|$, 
such that the strong coupling scale $\Lambda_{\ast}$ is defined as 
\begin{equation}
    \frac{\mathcal{L}_{3}}{\mathcal{L}_{2}}\sim\left(\frac{E_{\rm inf}}{\Lambda_{\ast}}\right)^{2}=\left(\frac{H}{\Lambda_{\ast}}\right)^{2},
\end{equation}
where we have assumed that the characteristic energy of inflation is given by $E_{\rm inf}=H$. In order to estimate $\Lambda_{\ast}$ we need to estimate $\mathcal{L}_{3}$ and $\mathcal{L}_{2}$. We use a set of heuristic rules
\begin{align}
    &\frac{\mathcal{L}_{3}}{\mathcal{L}_{3}}\sim \fnl\Delta_{\zeta} \,,\quad \pi\sim \frac{\Delta_{\zeta}}{H},\,\quad \dot{\pi}\sim H\pi\,,\quad \partial_{i}\pi \sim k \pi . \label{eq:apprulesI}
\end{align}
This assumes that time and spatial derivatives are evaluated at horizon crossing, as that is when most of the signal is imprinted, such that $ aH\sim c_{s}k$. The rules in \eqref{eq:apprulesI} generate
\begin{equation}
    \fnl\Delta_{\zeta}\sim \left(\frac{H}{\Lambda_{\ast}}\right)^{2}\quad \rightarrow \quad\fnl\sim\left(\frac{\fpi}{\Lambda_{\ast}}\right)^{2}\,,
\end{equation}
where we used Eq.~\ref{eq:PkEFToI} in the last equality. Now using the last two heuristic rules \eqref{eq:apprulesI} to find $\fnl$ in terms of $c_{s}$, we have
\begin{equation}
    \fnl\Delta_{\zeta}\sim\frac{\mathcal{L}_{3}}{\mathcal{L}_{2}}=\frac{\left(c_{s}^{2}-1\right)\dot{\pi}\left(\partial_{i}\pi\right)^{2}}{c_{s}^{2}\left(\partial_{i}\pi\right)^{2}}=\frac{c_{s}^{2}-1}{c_{s}^{2}}\Delta_{\zeta}\quad \rightarrow\quad \fnl\sim\frac{1}{c_{s}^{2}},
\end{equation}
which means that the strong coupling scale is given by $\Lambda_{\ast}\sim c_{s}\fpi$. Therefore, there is a lower bound for $c_{s}$ in order to avoid $H\sim\Lambda_{\ast}$ which is 
\begin{equation}
    H\ll \Lambda_{\ast}\quad \rightarrow\quad \sqrt{\Delta}_{\zeta}<c_{s}\,.
\end{equation}
This means that whenever $c_{s}\sim \sqrt{\Delta}_{\zeta}$, it should be expected to find a strongly coupled regime. In \cite{Baumann:2011su} it was shown that there are example where there are weakly coupled UV models that solve this strong coupling problem. 

\subsection{The Open Effective Field Theory of Inflation}

The approach of \cite{Salcedo:2024smn} departs from the EFToI in the central object of its construction. In the dissipative theory, the fundamental quantity is not an action $S$, but an influence functional $\mathcal{I}$. This is because observables such as the bispectrum are in-in quantities computed using the Schwinger–Keldysh (SK) formalism. Unlike scattering amplitudes, the SK path integral contains two time branches \cite{Liu:2018kfw,Glorioso:2017fpd,Glorioso:2016gsa}. Accordingly, the field is doubled, $\pi \to \pi_\pm$, with each copy living on a different branch of the contour. In the absence of dissipative dynamics, the influence functional reduces to the difference between the action evaluated on the two branches of the path integral
\begin{equation}
    \mathcal{I}\left[\pi_{+},\pi_{-}\right]=S\left[\pi_{+}\right]-S\left[\pi_{-}\right]\,.
\end{equation}
However, in the presence of dissipative dynamics there are interactions that are represented by cross-branch terms. These are denoted $\mathfrak{I}\left[\pi_{+},\pi_{-}\right]$
\begin{equation}
    \mathcal{I}\left[\pi_{+},\pi_{-}\right]=S\left[\pi_{+}\right]-S\left[\pi_{-}\right]+\mathfrak{I}\left[\pi_{+},\pi_{-}\right]\,.
\end{equation}
Like in the EFToI \cite{Cheung:2007st}, symmetries and fundamental principles are instrumental to find the appropriate form of the influence functional. The main principle for influence functional is the existence of a UV completion evolving unitarily \cite{Liu:2018kfw,Glorioso:2017fpd,Glorioso:2016gsa}, under an action $S$ depending on both the system variables and the degrees of freedom that compose the environment. The existence of this action leads to the non-equilibrium constraints
\begin{align}
    \mathcal{I}\left[\pi,\pi\right]&=0\,,\nonumber\\
    \mathcal{I}\left[\pi_{+},\pi_{-}\right]&=-\mathcal{I}^{*}\left[\pi_{+},\pi_{-}\right]\,,\label{eq:NEQconstraints}\\
    \text{Im}\left\{\mathcal{I}\left[\pi_{+},\pi_{-}\right]\right\}&\geq0\,.\nonumber
\end{align}
These conditions are more easily expressed in the Keldysh basis
\begin{equation}
    \pir=\frac{1}{2}\left(\pi_{+}+\pi_{-}\right)\,,\quad\pia=\pi_{+}-\pi_{-}.
\end{equation}
The first condition in \eqref{eq:NEQconstraints} implies that $\mathcal{I}$ is at least linear in $\pia$. The second condition in \eqref{eq:NEQconstraints} implies that terms (odd) even in $\pia$ have (real) imaginary coefficients. The third condition in \eqref{eq:NEQconstraints} implies that a linear combination of these imaginary coefficients is positive.

The symmetries of the action $S$ are inherited by the influence functional $\mathcal{I}$ in the form of the product group $G_{+}\times G_{-}$. However, the presence of dissipative dynamics breaks the product group into its diagonal component:
\begin{equation}\label{eq:coset}
    G_{+}\times G_{-}\rightarrow\text{Diag}\left[G_{+}\times G_{-}\right]\,,
\end{equation}
leading to a coset construction for the SK path integral \cite{Akyuz:2023lsm, Firat:2025upx}. In particular, if one has a non-linearly realised symmetry like in \eqref{eq:NLRtime}, then an influence functional that describes dissipative dynamics will only be invariant under the diagonal transformation
\begin{align}
    \pir(t,\textbf{x})&\rightarrow \pir(t+\xi,\textbf{x})-\xi\,,\qquad
    \pia\left(t,\textbf{x}\right)&\rightarrow\pia\left(t+\xi,\textbf{x}\right)\,. \label{eq:diagonalNLR}
\end{align}

\subsubsection{Power spectrum} 

Following \cite{Cheung:2007st}, the influence functional in \cite{Salcedo:2024smn} is a sum over invariant blocks under the transformation \eqref{eq:diagonalNLR} and that follows the constraints in \eqref{eq:NEQconstraints}. Upon expanding to quadratic order in $\pi$ and canonically normalizing, one obtain
\begin{equation}
    \mathcal{I}^{\left(2\right)}\left[\pir,\pia\right]=\int \dd^4x \sqrt{-g} \;S^{\left(2\right)},
\end{equation}
with
\begin{equation}\label{eq:appS2OEFToI}
    S^{\left(2\right)}=\dpir\dpia-\frac{c_{s}^{2}}{a^{2}\left(t\right)}\partial_{i}\pir\partial_{i}\pia-\gamma\dpir\pia+i\beta\pia^{2}.
\end{equation}
The first two terms correspond to the same as in the EFToI \eqref{eq:quadraticdec}. The third term corresponds to dissipation into the environment. The last term corresponds to the sourcing of the field $\pi$ by the environment. It is possible to translate the dynamics encoded by an influence functional into a stochastic Langevin equation for $\pi$
\begin{equation}
    \Ddot{\pi}+\left(3H+\gamma\right)\dot{\pi}-\frac{c_{s}^{2}}{a^{2}\left(t\right)}\partial^{2}\pi=\xi,
\end{equation}
where $\xi$ is an stochastic field that models the environment as a Gaussian noise, whose amplitude is fixed by $\beta$ \cite{Hubbard:1959ub,Stratonovich1957}. In the presence of dissipative dynamics, the power spectrum amplitude $P_{\zeta}$ is not just a function of $c_{s},H$ and $\fpi$. Now it also depends on $\beta$ and $\gamma$ through
\begin{equation}\label{eq:OEFToIpowerspectrum}
    \langle \hat{\zeta}_{\bfk_{1}}\hat{\zeta}_{\bfk_{2}}\rangle=\left(2\pi\right)^{3}\delta\left(\bfk_{1}+\bfk_{2}\right)\frac{2\pi^{2}}{k^{3}}\frac{\beta}{H^{2}}\frac{H^{4}}{f_{\pi}^{4}}2^{2\nu_\gamma}\frac{\Gamma(\nu_\gamma - 1)\Gamma(\nu_\gamma)^2}{\Gamma\!\left(\nu_\gamma - \tfrac{1}{2}\right)\Gamma\!\left(2\nu_\gamma - \tfrac{1}{2}\right)}\,,\quad \nu_{\gamma}=\frac{3}{2}+\frac{\gamma}{H}
\end{equation}
In general, dissipative dynamics enhance the scalar power spectrum while leaving the tensor power spectrum unaffected, which yields a very small tensor-to-scalar ratio $r$. 

\subsubsection{Dissipative bispectra} 

Dissipative dynamics lead to sixteen terms in the cubic order expansion of $\mathcal{I}$, compared to the two in \eqref{eq:EFToIcubic}. In this work we will only focus on the terms that are related to dissipation $\gamma$ and the speed of sound $c_{s}$ through the non-linearly realized symmetry \eqref{eq:diagonalNLR}. These are
\begin{align}\label{eq:OEFToIbispectrum}
    S^{(3)}=&\frac{c_{s}^{2}-1}{2\fpi^{2}}\left[\dpia\left(\partial_{i}\pir\right)^{2}+2\dpir\partial_{i}\pir\partial^{i}\pia-3\dpia\dpir^{2}\right]+\frac{\gamma}{2\fpi^{2}}\pia\left[\left(\partial_{i}\pir\right)^{2}-\dpir^{2}\right]+...
\end{align}
The first group of terms corresponds to self-interactions of the inflaton, analogous to those appearing in the EFToI \cite{Cheung:2007st}. The second group provides an example of non-linear dissipation. On top of these operators, dissipative dynamics generically introduces additional interactions, such as system-noise couplings (cubic vertices involving $\mathcal{O}(\pir\pia^2)$) and non-Gaussian noise terms (vertices cubic in $\pia$). These interactions are all present in the model of \cite{Creminelli:2023aly} and will be systematically studied in \cite{Longpaper}.

The bispectra generated by the vertices in \eqref{eq:OEFToIbispectrum} are parametrized by $\gamma$ and $c_s$. Their shapes are smooth and exhibit partial overlap with the equilateral \eqref{eq:eqShape} and orthogonal \eqref{eq:orthShape} templates \cite{Salcedo:2024smn}, implying an observable amplitude $\fnl^{\rm obs}\sim\mathcal{O}(50)$. Two phenomenologically distinct regimes arise.
In the low-dissipation limit, $\gamma \ll H$, the bispectrum peaks in the folded configuration $k_1 = 2k_2 = 2k_3$, where the momentum triangle is isosceles and flat. The enhancement originates from highly blue-shifted perturbations sourced by the environment that do not decay sufficiently early and therefore survive until reheating. These modes are insensitive to curvature effects and approximately conserve energy, which favours folded momentum configurations.
In the opposite regime of strong dissipation, $\gamma \gg H$, the dominant fluctuations reaching the reheating surface are produced shortly before the end of inflation. Their comparable wavelengths lead to a bispectrum close to the equilateral shape. In this work the dominant operator in this limit is $(\partial_i \pi_r)^2\pia $, whose shape overlaps with both the equilateral and orthogonal templates, resulting in an asymptotic value of $\fnl$ that differs from that of a purely equilateral bispectrum.

Thanks to the parametrization in \eqref{eq:definitionfnlshape}, the dependence on $\beta$ and $\fpi$ cancels between the bispectrum and the power spectrum. The bispectrum generated by the operators in \eqref{eq:OEFToIbispectrum} scales as $\beta^{2}$, while the power spectrum satisfies $P_{\zeta} \propto \beta$ \eqref{eq:OEFToIpowerspectrum}. Consequently, the entire $\beta$ dependence of $B$ is absorbed into $P_{\zeta}^{2}$. An analogous cancellation occurs for $\fpi$. As a result, both the shape and the amplitude of the bispectrum, $\fnl$, depend only on the dissipation parameter and the speed of sound.
\begin{equation}
    S=S\left(k_{1},k_{2},k_{3};\gamma,c_{s}^{2}\right)\quad,\quad \fnl=\fnl\left(\gamma,c_{s}^{2}\right)
\end{equation}
Therefore, studying the bispectra sourced by Eq.~\eqref{eq:OEFToIbispectrum} is akin to constraining the $\left(\gamma,c_{s}\right)$ space. A first approach would restrict to only consider the operators sourced by $\gamma$. This corresponds to exploring the line $c_{s}=1$. Through the heuristic estimates developed in \cite{LopezNacir:2011kk, Salcedo:2024smn} we can explain why the predicted $\fnl$ is too small if $c_{s}=1$ unless $\gamma\sim 200 H$. Expanding the parameter space to include $c_{s}$ allows for a larger $\fnl$ at smaller dissipation. Analysis then explores the whole $\left(\gamma,c_{s}\right)$ space, as done in the main text.  

\subsubsection{Computing dissipative bispectra} 

The computation of the bispectrum can be convoluted. Previous works \cite{LopezNacir:2011kk,Creminelli:2023aly} expand the Langevin equation to NLO. This requires solving the equation for $\pi$ to leading order in the couplings and perform a set of time ordered integrals. In \cite{Salcedo:2024smn} some of the authors of this Letter derived a set of Feynman rules to compute dissipative bispectrum and showed the equivalence to the Langevin equation approach from \cite{LopezNacir:2011kk,Creminelli:2023aly}. Here we present a review on how dissipative bispectra are derived. We start by considering the bispectrum in the Schwinger-Keldysh formalism
\begin{equation}
    \langle\hat{\pi}_{\bfk_{1}}\hat{\pi}_{\bfk_{2}}\hat{\pi}_{\bfk_{3}}\rangle=\int \dd\pi_{\bfk}\pi_{\bfk_{1}}\pi_{\bfk_{2}}\pi_{\bfk_{3}}\int_{\rm BD}^{\pi_{k}}\mathcal{D}\pi_{a}\int_{\rm BD}^{0}\mathcal{D}\pi_{a} e^{i\mathcal{I}\left[\pi_{r},\pi_{a}\right]}\;,
\end{equation}
where we perform a path integral over every possible field configuration of $\pi_{a}$ between the Bunch-Davies vacuum condition to the infinite past and a vanishing field configuration at the future boundary of the path integral. Indeed, at the future boundary, the values of both fields $\pi_{\pm}=\pi_{\bfk}$ are identified such that $\pia=\pi_{+}-\pi_{-} = 0$. The second path integral is over every possible field configuration of $\pi_{r}$ between the Bunch-Davies vacuum condition to the infinite past and the field profile $\pi_{\bfk}$ on the future boundary. These paths integrals provide a weight for each $\pi_{\bfk}$ configuration, over which we perform the final integration. Hence, we can take $\pi_{\bfk_{1}}\pi_{\bfk_{2}}\pi_{\bfk_{3}}$ into the $\pi_{r,a}$ path integrals such that 
\begin{equation}
    \langle\hat{\pi}_{\bfk_{1}}\hat{\pi}_{\bfk_{2}}\hat{\pi}_{\bfk_{3}}\rangle=\int \dd\pi_{\bfk}\int_{\rm BD}^{\pi_{k}}\mathcal{D}\pi_{a}\int_{\rm BD}^{0}\mathcal{D}\pi_{a} \pi_{r,\bfk_{1}}\left(t_{0}\right)\pi_{r,\bfk_{2}}\left(t_{0}\right)\pi_{r,\bfk_{3}}\left(t_{0}\right)e^{i\mathcal{I}\left[\pi_{r},\pi_{a}\right]}\,.
\end{equation}

Working in perturbation theory and considering only the leading order contribution to the bispectrum, we perform the split
\begin{equation}
    \mathcal{I}\left[\pi_{r},\pi_{a}\right]=\mathcal{I}^{(2)}\left[\pi_{r},\pi_{a}\right]+\lambda \mathcal{I}^{(3)}\left[\pi_{r},\pi_{a}\right],
\end{equation}
where $\mathcal{I}^{(2)}\left[\pi_{r},\pi_{a}\right]$ is the quadratic functional described in Eq.~\eqref{eq:appS2OEFToI} and $\mathcal{I}^{(3)}\left[\pi_{r},\pi_{a}\right]$ encompasses the cubic operators considered in \cite{Salcedo:2024smn}. At leading order in $\mathcal{O}\left(\lambda\right)$ we would find:
\begin{equation}
    \langle\hat{\pi}_{\bfk_{1}}\hat{\pi}_{\bfk_{2}}\hat{\pi}_{\bfk_{3}}\rangle=i\lambda\int d\pi_{\bfk}\int_{\rm BD}^{\pi_{k}}\mathcal{D}\pi_{a}\int_{\rm BD}^{0}\mathcal{D}\pi_{a} \left[\pi_{r,\bfk_{1}}\left(t_{0}\right)\pi_{r,\bfk_{2}}\left(t_{0}\right)\pi_{r,\bfk_{3}}\left(t_{0}\right)\mathcal{I}^{(3)}\left[\pi_{r},\pi_{a}\right]\right]e^{i\mathcal{I}^{(2)}\left[\pi_{r},\pi_{a}\right]},
\end{equation}
which has reduced the computation of the bispectrum in the interacting theory to computing the expectation value of $\pi^{3}\, \mathcal{I}^{(3)}$ in the Gaussian theory. This is done through Wick contractions, of which there will be two types
\begin{align}
    \langle\pi_{r}(t_{1})\pi_{r}(t_{2})\rangle=-iG^{K}\left(t_{1},t_{2}\right)\,,\qquad  \langle\pi_{r}(t_{1})\pi_{a}(t_{2})\rangle=G^{R}\left(t_{1},t_{2}\right)\,.
\end{align}
In practice, $t_{1}$ is taking to the future boundary of the Schwinger-Keldysh path integral and $t_{2}$ corresponds to the time of the single interaction vertex under consideration. $G^{K}\left(t_{1},t_{2}\right)$ is the \textit{Keldysh propagator} and $G^{R}\left(t_{1},t_{2}\right)$ is the \textit{retarded propagator}. The latter describes the propagation of fluctuations of the inflaton affected by the dissipation into the environment while the former describes the propagation of fluctuations sourced by the noise from the environment. These are computed by considering the partition function of the quadratic theory \cite{Salcedo:2024smn} and are the solution Green's equations built from differential operators determined by the quadratic action \eqref{eq:appS2OEFToI}:
\begin{align}
   \hat{D}_R= &a^2 \partial_\eta^2 
+ a^3 \left(\frac{2H}{a} + \gamma \right)\partial_\eta 
- a^2 c_s^2 \partial_i^2\,,\quad \hat{D}_{K}(x,z)=a^{4}\beta\,,
\end{align}
such that
\begin{align}
    \hat{D}_R(x) G^{R}(x,y)&=-i\delta\left(x-y\right)\,,\\
    \hat{D}_R(x) G^{K}(x,y)&=2i\hat{D}_{K}(x)G^{R}(z,y)\,.
\end{align}
Solving these equations yields the propagators in \eqref{eq:GRB2b} and \eqref{eq:GK}. 

\subsubsection{Strong coupling scales} 

In a similar manner to the previous section we can consider the regime for which perturbation theory breaks down, leading to an strong coupled description. This can happen when $\fnl\gg1$. Therefore, we assume we find ourselves in the large dissipation and small speed of sound regime. We leave the limit $\gamma\leq H$ and $c_{s}\ll1$ for future study. To derive the strong coupling scale we start from the relation:
\begin{equation}
    \fnl\sim\left(\frac{\fpi}{\Lambda_{\ast}}\right)^{2},
\end{equation}
which can be inverted to 
\begin{equation}
    \Lambda_{\ast}^{2}\sim\frac{\fpi^{2}}{\fnl}\sim 4c_{s}^{2}\frac{\gamma}{H}\fpi^{2}\,,
\end{equation}
using the main text heuristic estimate \eqref{eq:fnltheo}. Similarly to the EFToI case, we can derive a bound for $c_{s}$ and $\gamma$ by considering only the weakly coupled regime of the theory, such that
\begin{equation}
    E_{\rm sys}\ll \Lambda_{\ast}\qquad \rightarrow \qquad  \frac{1}{4c_{s}^{2}}\frac{\gamma}{H}\ll \sqrt{\Delta_{\zeta}}\,.
\end{equation}
We conclude that the theory only becomes strongly coupled very far away from the region preferred by data. 

%%%%%%%%%%%%%%%%%%%%%%%%%%%%%%%%%%%%%%%%%%%%%%%%%%%%%%%%%%

\subsection{Primordial bispectrum constraints from CMB data}

In the weak non-Gaussian limit, we model the probability of observing the bispectrum $B_{\boldsymbol{\ell}}^{\rm (obs.)}$ in the sky, given the inflation theory $B_{\boldsymbol{\ell}}^{\rm (th.)}$, as a multivariate normal distribution. Using the condensed notation: 
\begin{equation}
    \boldsymbol{\ell}_j  \coloneqq (\ell_j,m_j) ,\quad
    L \coloneqq (\ell_1,\ell_2,\ell_3) ,\quad
    \boldsymbol{L} \coloneqq (\boldsymbol{\ell_1},\boldsymbol{\ell_2},\boldsymbol{\ell_3}), 
\end{equation}
we can write the likelihood as
\begin{equation}\label{eqn:likelihood_appendix}
    \L = \dfrac{1}{\sqrt{(2\pi)^N \det \Sigma}} \exp \left(-\dfrac{1}{2} \sum_{\boldsymbol{\ell_i},\boldsymbol{\ell_j}} \left(B_{\boldsymbol{\ell_i}}^{\rm (obs.)} -  B_{\boldsymbol{\ell_i}}^{\rm (th.)}\right) \Sigma^{-1}_{ij} \left( B_{\boldsymbol{\ell_j}}^{\rm (obs.)} -  B_{\boldsymbol{\ell_j}}^{\rm (th.)}\right) \right).
\end{equation}
Hence, the observed data acts as the random variable centered around mean $\mu = B_{\boldsymbol{\ell}}^{\rm (th.)}$, with the covariance matrix $\Sigma$.

The variables are further modified so that the resulting covariance matrix reduces to identity while keeping the estimator unbiased.
Assuming $a_{\ell m}$'s to be weakly non-Gaussian and noting that cosmic variance dominates over measurement errors, the variables become
\begin{equation}\label{eqn: B tilde}
        \Tilde{B}_{\boldsymbol{L}}^{\mathrm{obs}} = \dfrac{a_{\boldsymbol{\ell_1}}a_{\boldsymbol{\ell_2}} a_{\boldsymbol{\ell_3}} - \langle a_{\boldsymbol{\ell_1}}a_{\boldsymbol{\ell_2}} \rangle a_{\boldsymbol{\ell_3}}
    - \langle a_{\boldsymbol{\ell_2}}a_{\boldsymbol{\ell_3}} \rangle a_{\boldsymbol{\ell_1}} -
    \langle a_{\boldsymbol{\ell_1}}a_{\boldsymbol{\ell_3}} \rangle a_{\boldsymbol{\ell_2}}}{\sqrt{\Delta_{\boldsymbol{\ell_1}\boldsymbol{\ell_2}\boldsymbol{\ell_3}}C_{\ell_1}C_{\ell_2}C_{\ell_3}}}
    \quad
    \text{and}
    \quad
    \Tilde{B}_{\boldsymbol{L}}^{(\rm th)} = \dfrac{ f_{\rm NL} \mathcal{G}_{\boldsymbol{L}}B_{L}^{(\rm th)}}{\sqrt{\Delta_{\boldsymbol{\ell_1}\boldsymbol{\ell_2}\boldsymbol{\ell_3}}C_{\ell_1}C_{\ell_2}C_{\ell_3}}},
\end{equation}
where $\Delta_{\boldsymbol{\ell_1}\boldsymbol{\ell_2}\boldsymbol{\ell_3}}$ is a symmetry factor equal to 6 (if $\boldsymbol{\ell_1} =\boldsymbol{\ell_2} = \boldsymbol{\ell_3}$), equal to 2 (if two $\boldsymbol{\ell}$'s are identical) or equal to 1 (if all three are different). Theoretically, we would define $C_{\boldsymbol{\ell_1\ell_2}} = \langle a_{\boldsymbol{\ell_1}} a_{\boldsymbol{\ell_2}} \rangle$, but in practice we either take $\langle a_{\boldsymbol{\ell_1}} a_{\boldsymbol{\ell_2}} \rangle_{\textit{MC}}$ or approximate $C_{\boldsymbol{\ell_1\ell_2}} \approx C_\ell \delta_{\boldsymbol{\ell_1\ell_2}}$. With these variables, the error in each $\boldsymbol{L}$-mode is independent with unit variance.

With these modifications, the log likelihood \eqref{eqn:likelihood_appendix} simplifies to
\begin{equation}\label{eqn:-2logLv2}
    -2 \log \L = \text{Const.} +  \sum_{\boldsymbol{\ell_i}} \left(\tilde B_{\boldsymbol{\ell_i}}^{\rm (obs.)} - \tilde B_{\boldsymbol{\ell_i}}^{\rm (th)} \right)^2 = \text{Const.} - 2 f_{\rm NL} P + f_{\rm NL}^2 F,
\end{equation}
where the constant is the data term term and does not depend on theory, and we simplified using $F$ and $P$ as
\begin{equation}
\begin{split}
    F &\equiv \tilde{B}^{\rm th^2}
    = 
    \sum_{\boldsymbol{\ell_1} \leq \boldsymbol{\ell_2} \leq \boldsymbol{\ell_3}} \dfrac{\mathcal{G}_{\boldsymbol{L}}^2 b_L^{(\rm th)^2}}{\Delta_{\boldsymbol{\ell_1}\boldsymbol{\ell_2}\boldsymbol{\ell_3}}C_{\ell_1}C_{\ell_2}C_{\ell_3}}
    =
    \sum_{\boldsymbol{\ell_1} \boldsymbol{\ell_2} \boldsymbol{\ell_3}} \dfrac{\mathcal{G}_{\boldsymbol{L}}^2 b_L^{(\rm th)^2}}{6C_{\ell_1}C_{\ell_2}C_{\ell_3}}
    = 
    \sum_{\ell_1 \ell_2 \ell_3} \dfrac{h_{L}^2 b_L^{(\rm th)^2}}{6C_{\ell_1}C_{\ell_2}C_{\ell_3}}\\
    P &\equiv \tilde{B}^{\rm th} \cdot \tilde{B}^{\mathrm{obs}} 
    = 
    \sum_{\boldsymbol{\ell_1} \boldsymbol{\ell_2} \boldsymbol{\ell_3}} \dfrac{\mathcal{G}_{\boldsymbol{L}} b_L^{(\rm th)}}{6C_{\ell_1}C_{\ell_2}C_{\ell_3}}
    [a_{\boldsymbol{\ell_1}}a_{\boldsymbol{\ell_2}} a_{\boldsymbol{\ell_3}} - \langle a_{\boldsymbol{\ell_1}}a_{\boldsymbol{\ell_2}} \rangle a_{\boldsymbol{\ell_3}}
    - \langle a_{\boldsymbol{\ell_2}}a_{\boldsymbol{\ell_3}} \rangle a_{\boldsymbol{\ell_1}} -
    \langle a_{\boldsymbol{\ell_1}}a_{\boldsymbol{\ell_3}} \rangle a_{\boldsymbol{\ell_2}}].
\end{split}
\end{equation}
Modal decomposition turns the $\mathcal{O}(\ell_{\rm max}^5)$ summation in $F$ and $P$ into a scalar product (assuming orthonormal basis) between the theoretical and the observed bispectrum modes, $(\alpha_i)$ and $(\beta_i)$, respectively:
\begin{equation}
\begin{split}
    F = \dfrac{1}{6}\sum_{i=0}^{n} \alpha_i^2 \quad \rm {and} \quad
    \textit{P} = \dfrac{1}{6} \sum_{i=0}^{n} \alpha_i\beta_i.
\end{split}
\end{equation}
The goal of using a bispectrum estimator is to test the statistical significance of the data in supporting the theory. As written in Eq.~\eqref{eqn: B tilde}, the theoretical bispectrum has a certain shape ($B_{\ell_1,\ell_2, \ell_3}$) that encodes the dynamics of quantum interactions in the early Universe, and an overall amplitude of the fluctuations $f_{\rm NL}$. Depending on how strong the theoretical predictions can be, we identify two possible cases:
\begin{enumerate}
    \item The bispectrum \textit{shape} is uniquely determined by the theory, but its \textit{amplitude} is arbitrary up to a constant prefactor. Such theory can successfuly capture the kinematics of inflation fields, but is agnostic to its magnitudes.
    \item Theoretical calculations give unique signature of inflation, thus unambiguously determining the bispectrum's shape and its amplitude.
\end{enumerate}
\paragraph{1. Arbitrary amplitude} In the former case, since the theoretical shape admits any amplitude, we look for one which maximizes the likelihood. By varying $f_{\rm NL}$ in Eq. \eqref{eqn:-2logLv2}, the maximum likelihood estimator (MLE) reduces to finding the best-fit $f_{\rm NL}$
\begin{equation}
    \hat f_{\rm NL} = \dfrac{P}{F} = \dfrac{\sum_i \alpha_i \beta_i}{\sum_i \alpha_i^2}.
\end{equation}
The corresponding Fisher error is given by $\hat \sigma_{f_{\rm  NL}} = \sqrt{F^{-1}} = \sqrt{ 6/\sum_i\alpha_i^2}$, but a more realistic error estimate $\sigma_{f_{\rm  NL}}$ is obtained using simulated Gaussian CMB maps. The quoted constraints on the non-Gaussianities from such models are given as $\hat f_{\rm NL} \pm \hat \sigma_{f_{\rm  NL}}$.

However, note that such a result depends on the overall scaling of the bispectrum shape. If we scale the bispectrum amplitude by a factor of $A$, one would obtain $f_{\rm NL}$ constraint of the form $A^{-1}\hat f_{\rm NL} \pm A^{-1}\sigma_{f_{\rm  NL}}$. Hence, the useful and amplitude-independent quantity is the signal-to-noise ratio (SNR):
\begin{equation}
    \tilde \sigma_{\rm SNR} = \dfrac{\hat f_{\rm NL}}{ \sigma_{f_{\rm  NL}}}.
\end{equation}
\paragraph{2. Fixed amplitude} In the latter case, the $f_{\rm NL}$ parameter is fixed by the theory. Such templates can be directly compared with the data in Eq. \eqref{eqn:-2logLv2}:
\begin{equation}
    -2\log \L \propto \dfrac{1}{6}\left(f_{\rm NL}^2 \sum_{i=0}^{n} \alpha_i^2 - 2 f_{\rm NL} \sum_{i=0}^{n} \alpha_i \beta_i\right).
\end{equation}
If the theory involves a range of free parameters $\boldsymbol{\theta} = (\theta_1, \theta_2, ...)$ which affect the bispectrum, we can produce a posterior distribution $\mathcal{P}$, by applying a Bayesian prior $\Pi(\boldsymbol{\theta})$
\begin{equation}
    \mathcal{P}(\boldsymbol{\theta}) \propto \L(\boldsymbol{\theta} | \rm{data}) \cdot \Pi(\boldsymbol{\theta}).
\end{equation}
With the posterior appropriately normalized, one can rule out certain regions of the parameter space at some significance level.\\

\end{widetext}
\end{document}